# Optimally efficient, high power, Yb:fiber laser pumped, near- to mid-infrared picosecond optical parametric oscillator

**Omid Kokabee**
*Ward 350 of Evin Prison, Tehran, Iran*
*kokabee@gmail.com*

**Abstract:** Absolute output power optimization and performance of a near- to mid-infrared picosecond optical parametric oscillator (OPO) is studied at two high pump powers using a widely-tunable output coupling (OC) technique which provides 15% to 68% OC. The MgO:PPLN-based OPO is synchronously pumped at 81.1 MHz by an Yb:fiber laser with double-peak spectrum. At 2 W pump power, maximum signal (at 1.46 μm) and idler (at 3.92 μm) power of 670 mW and 270 mW are obtained at 27% OC at 47% total extraction efficiency and 58% pump power depletion where at 15.5 W pump power, 7.4 W signal and 2.7 W idler power are extracted at 53% OC at 65% total extraction efficiency and 80% pump depletion. With respect to non-optimum points, OPO provides signal pulses with norrower single-peak spectrum, smaller time-bandwidth product, much better circular single-mode $TEM_{00}$ spatial profile and passive peak-to-peak power stability of ±4.6% at 2 W and ±1% at 15.5 W pump power in optimum power extraction points.

## 1. Introduction

High-power synchronously pumped optical parametric oscillators (SPOPOs) are attractive sources of high-repetition-rate, wavelength-tunable femtosecond (fs) and picosecond (ps) pulses from the UV to mid-IR, used in a wide variety of applications, including laser-induced cooling [1], Coherent Anti-Stokes Raman Scattering (CARS) microscopy [2], non-invasive microscopy [3] and telecommunications [4]. Among nonlinear crystals, periodically-poled $LiNbO_3$ (PPLN) is known as the most effective material for ultrashort high-power operation in the near- to mid-infrared with a broad transparency range from 0.35 to 6.8 μm [5-7]. It is also capable of fulfilling the requirements of high gain due to its large nonlinearity, wide tunability due to flexible noncritical phase-matching across its transparency range and long interaction length. Nevertheless, the main barrier in the application of PPLN for light generation in nonlinear optical devices is its relatively low optical damage resistance when exposed to increasing levels of visible light generated through non-phase-matched or higher order phase-matching processes in comparison to other competing oxide crystals such as $BaB_2O_4$ (BBO) and $LiTaO_3$ (LT). This damage can deteriorate the temporal, spatial and power stability of SPOPO output, limiting efficient and practical operation. PPLN is doped with MgO to increase the photorefractive damage threshold by at least three orders of magnitude and to avoid beam distortion and green-induced infrared absorption as well as enhancing the operation in mid-infrared range [8-10].

A variety of bulky mode-locked laser sources has been used to pump ultrashort SPOPOs including diode laser MOPA system [11], fiber-amplified gain-switched laser diode [12], Ti:sapphire laser [6,10], Nd:YAG laser [13], diode-pumped $Nd:YVO_4$ laser [4], Nd:YLF laser [14] and Yb:KGW slab laser [15]. However, due to rapid developments in high-power fiber lasers, noticeable interest has grown in pumping SPOPOs by Er-doped [16,17] and Yb-doped [7, 18-20] fiber lasers in recent years. Fiber lasers are particularly advantageous for OPO pumping when compared to solid-state lasers, on account of their excellent beam quality, high stability and ultrahigh electrical-to-optical conversion efficiency which are essential for highly efficient frequency conversion that can be maintained in a more simplified, power scalable, compact, portable and cost-effective way without recourse to complicated thermal management schemes.

Additionally, high-power operation involves complications inside the cavity and gain medium which can have a direct effect on the efficiency, stability and temporal, spatial and spectral quality of SPOPO output. In an SPOPO which is pumped with a high quality fiber laser, appropriate power extraction is vital to achieve most favorable results with highest obtainable efficiency. The most practical method to extract the maximum power from an OPO requires experimental evaluation of different output couplers. However, absolute optimization of output power requires deployment of numerous mirrors of discrete transmission values which makes its development very difficult and expensive. Moreover, any variations in operating conditions, such as the pumping level, change the optimum output coupling, requiring different mirrors. Recently, a simple and powerful technique was introduced for absolute optimization of output power from optical oscillators using an anti-resonant ring interferometer (ARR) [21]. Using this technique, this work reports study and comparison of absolute power optimization and spectral, temporal, spatial and power stability performance of an Yb:fiber laser pumped MgO:PPLN-based ps SPOPO at a select signal wavelength at two high pump powers of 2 W and 15.5 W.

**2. SPOPO experimental setup**

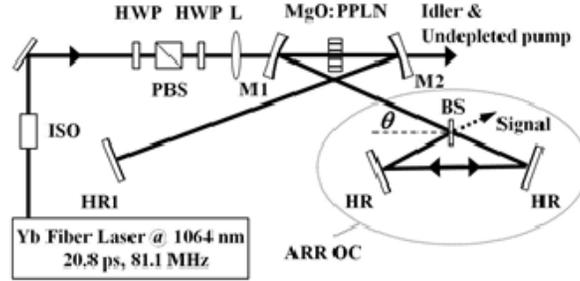

Fig. 1. Schematic of the ps SPOPO experimental layout with an antiresonat-ring interferometer output coupler (ARR OC) in one arm

The passively mode-locked ps Yb:fiber laser (Fianium, FemtoPower FP1060-20) delivers up to 20 W of average power at 81.1 MHz repetition-rate at 1.064 μm providing 20.8 ps (FWHM) pulses with a double-peak spectrum in a measured bandwidth of 1.38 nm (FWHM). As shown in Fig. 1, after transmission through optical isolator (ISO), linear polarization of pump beam is restored by a λ/2 waveplate (HWP) and a polarizing beamsplitter (PBS) which are also used for pump power attenuation. Then using a two lens telescope (not shown here), another HWP (used to rotate the pump beam polarization) and a focusing lens, pump beam is focused into a 50-mm-long and 1-mm-thick 5 mol.% CLN-type congruent MgO:PPLN crystal (HC Photonics, Taiwan), which contains five 1.2-mm-wide parallel gratings, equally spaced in period from 28.5 μm to 30.5 μm. According to the manufacturer's data sheet, this crystal has a $d_{33}$ nonlinear coefficient of 25 pm/V with a transparency range of 0.33 to 5.5 μm. In high-power regime due to the existence of very high intracavity powers, increased beam absorption leads to increased crystal temperature which is not favorable for effective operation. To maintain the crystal temperature with great accuracy, crystal is placed in a temperature-controlled home-made oven with a temperature stability of ±0.1 °C, adjustable from room temperature to 200 °C. The pump is focused to a beam waist radius of ~45 μm at the center of the crystal, resulting in a focusing parameter of ξ~1.94. The crystal faces are antireflection-coated for signal ($R<1\%$ over 1.45-1.75 μm), with high transmission for pump ($T>97\%$) and idler ($T>95.5\%$ over 3-4.2 μm). Wavelength tuning is achieved by varying the crystal temperature for different grating periods, resulting in 205 nm signal range over 1.43–1.63 μm with 1100 nm idler range over 4.16–3.06 μm.

The SPOPO consists of a four-mirror standing-wave cavity configuration (Fig. 1), comprising two curved mirrors (M1 and M2, $r= 20$ cm, $CaF_2$ substrate) and one plane end mirror (HR1), which are highly reflecting for the signal ($R>99.9\%$ over 1.4-1.7 μm) and highly transmitting for the idler ($T>87\%$ over 3-4.2 μm) and pump ($T\sim92\%$). In the other end, instead of a conventional output coupler, we place an ARR interferometer (ARR OC). Two plane mirrors (HR) and their coatings are similar to HR1. The pellicle beam splitter (BS) is a 2-μm-thick nitrocellulose membrane (Thorlabs BP145B3) coated for 1-2 μm range. ARR's transmission and reflection can be tuned by changing the angle θ through rotating the BS. BS is mounted on a finely adjustable rotational stage rotating around an axis normal to the page. BS is selected to be simple, inexpensive, readily available, and thin to avoid pulse broadening due to the dispersion. The signal output is extracted from one side of BS and the idler radiation and undepleted pump are measured after M2 after applying appropriate special filters. The cavity configuration resulted in a signal beam waist inside the crystal that provides optimum overlap with the pump beam ($b_p \approx b_s$).

**3. Experimental results**

As shown in Fig. 2(a), due to change in BS's transmission and reflection by changing θ [21] from 10° to 60°, ARR OC can provide continuous output coupling values from 15% to 68% in the SPOPO operation wavelength of 1.46 μm. However OC beyond this range is not needed for this study, it is only limited physically by beam and BS projected cross-section size normal to the beam.

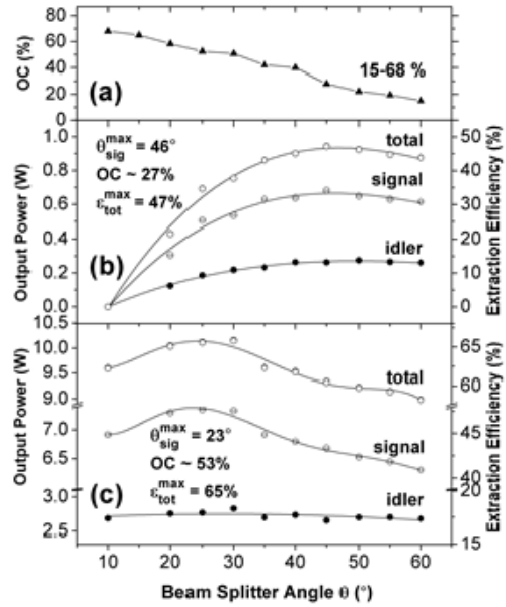

Fig. 2. (a) Experimental values of output-coupling (%OC) versus BS angle θ. Signal, idler and total extracted powers and extraction efficiency of ps SPOPO through ARR OC versus BS angle θ at (b) 2W pump power and (c) 15.5 W pump power.

The search for absolute SPOPO power output coupling at signal wavelength of 1.46 μm and its corresponding idler at 3.92 μm is carried out at two different high pump powers by recording highest extractable power at the whole available OC range . Maximum available pump power at the input to the crystal is 15.5 W due to overall loss through the optical isolator and transmission and focusing optics. Figure 2(b) and 2(c) show the average signal, idler and total power and the corresponding total extraction efficiency across BS angle at 2 W and 15.5 W of pump power at the input to the crystal, respectively. At 2 W pump power, SPOPO is not operational at θ=10° which corresponds to very high OC of 68%. By increasing the BS angle which corresponds to reducing OC value from high figures, SPOPO starts working and extractable power increases rapidly. This phenomenon is understandable since high OC drops the internal power severely and results in turning off the SPOPO or the reduction of conversion efficiency. Later, maximum signal output power of 670 mW and idler output power of 270 mW with total extraction efficiency of 47% occur at θ=46° corresponding to ~27% OC and 58% pump power depletion. Beyond θ=46°, for OC values less than 27%, by intracavity power buildup output power gradually drops which could be due to increased degenerating nonlinear effects which reduces the conversion efficiency. For 15.5 W pump power, since pump power is extremely high, SPOPO operates at very high power at all angles and OC values. The maximum signal and total output power of 7.4 W and 10. 1 W at 48% and 65% extraction efficiencies, respectively, happen at θ=23° corresponding to ~53% OC with about 80% pump power depletion. As can be found from Fig. 2(c) idler power remains almost the same at around 2.7 W across the whole OC range and output power optimization happens for the signal. According to Manley-Rowe relations expected idler power for 7.4 W signal is 2.74 W which is a bit higher than experimental value. This idler power loss is expected since the transmission of crystal coating is around 96% at idler wavelength.

Spectral characteristics of signal pulses are studied at three different angles, at optimum OC and two non-optimum OC, for each pump power. Upper and lower rows in Fig. 3 correspond to 2 W and 15.5 W pump power, respectively. In upper row, θ=23°, 46° and 60° correspond to 53% , 27% and 15% OC, respectively where in lower row, θ=10°, 23° and 46° correspond to 68%, 53% and 27% OC, respectively. In each row, the middle profile corresponds to pulses at the optimum OC point.

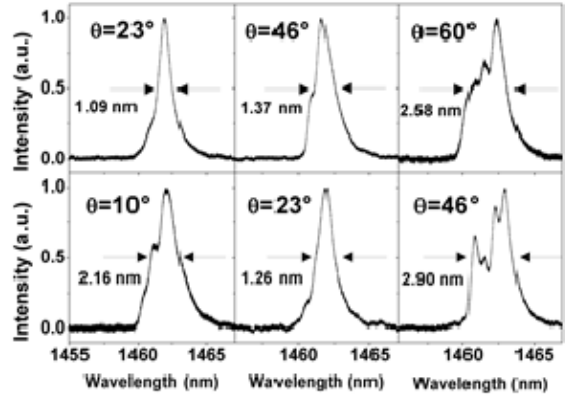

Fig. 3. Spectra of SPOPO output signal at three different angles at 2 W (upper row) and 15.5 W pump power (lower row). The middle one corresponds to the optimum OC.

For 2 W pump power, FWHM of optimum spectrum is 1.37 nm and for 15.5 W is 1.26 nm, both close to 1.38 nm pump pulses spectral bandwidth but in single-peak profile which is of high significance for the required applications. At both pump powers, at large angles (or small OC) due to high internal signal power, nonlinear effects deteriorate the spectrum making it broader with more peaks. For the smaller angles (or higher OC) at 2 W pump power, at θ=23°, OC is so high that gain is only limited to the central high-power part of the spectrum and leads to final narrower spectrum than optimum point. But at 15.5 W, since pump power is very high, interplay between different nonlinear effects which get stronger with getting farther from optimum point results in broader and double peak spectrum. It should be noted that at 15.5 W pump power, crystal temperature should be lowered by 9 °C to have the similar wavelength as for pump power at 2 W which could correspond to refraction index modification due to considerable increase in pump intensity in the crystal. Additionally, there is no evidence of power saturation which implies further room for pump power scaling.

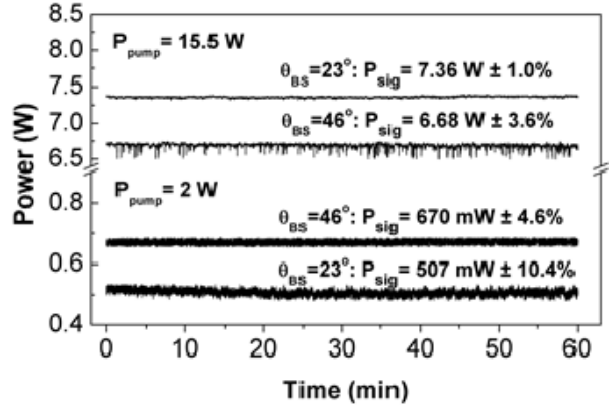

Fig. 4. Comparisons of signal output power stability of ps SPOPO at optimum and one select non-optimum BS angle for two different pump powers of 15.5 W (upper couple) and 2 W (lower couple).

Figure 4 shows the output power stability of signal at 1.46 μm at 2 W and 15.5 W of pump power at two different angles (output coupling values) one corresponding to optimum output coupling point. The MgO:PPLN crystal and the concave mirrors (M1 and M2) are isolated from laboratory air by enclosure in a Perspex box. Over an hour, the signal exhibited passive peak-to-peak power stability of ±10.4% at 23° and ±4.6% at 46° (optimum angle) for 2 W pump power and ±3.6% at 46° and ±1% at 23° (optimum angle) for 15.5 W pump power. In both pump powers, signal power stability in optimum OC points is significantly better than non-optimum points.

Temporal measurements of signal pulses are performed using two-photon intensity autocorrelation and in all cases SPOPO delivered pulses with de-convoluted FWHM pulse duration around 18 ps (assuming a sech$^2$ pulse shape). Resulting time-bandwidth product ($\Delta\tau\,\Delta\upsilon$) of signal at optimum OC at both pump powers (~3.5 at 2 W and ~3.2 at 15.5 W) are comparable to time-bandwidth product of pump (~7.6) even with better single-peak spectra which implies that SPOPO follows pump in temporal and spectral scales with favorable pulse quality enhancement in frequency conversion.

Finally, the spatial profile of the signal beams is measured at the same points as we measured the spectra and is shown in Fig. 5 with the same order. All points are characterized by single-mode TEM$_{00}$ spatial profile with best circular ones at the optimum points which are the middle ones in both rows.

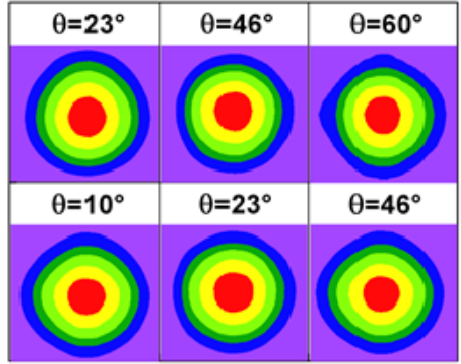

Fig. 5. Spatial beam profile of signal output at three different angles at 2 W (upper row) and 15.5 W pump power (lower row). The middle one corresponds to the optimum OC.

## 4. Conclusions

In conclusion, using a recently introduced widely-tunable output coupling (OC) technique based on antiresonant-ring (ARR) interferometer, an Yb:fiber laser pumped high repetition rate near- to mid-infrared MgO:PPLN ps optical parametric oscillator (OPO) is studied for absolute output power optimization in two different high pump powers. At 2 W pump power, maximum signal power extraction occurs at 27% OC at 47% total extraction efficiency where at 15.5 W pump power, it happens at 53% OC at 65% total extraction. These optimum points accompanied with considerable superior performance of OPO in spectrum, spatial profile and power stability demonstrates that these systems are capable of avoiding non-optimized OC and improper manipulation of internally resonating power which leads to very unstable and practically unacceptable operation. There is still room for further improvement in power extraction efficiency and performance of this SPOPO system. These methods at least include the use of a pump with better spectrum and still higher power; better crystal and optical elements coatings providing higher transmission for idler and pump and higher reflection for signal; intracavity dispersion compensation elements and finally, for enhanced stability, more stringent mechanical and thermal isolation of the system using better mirror mounts and oven.

## Acknowledgments

This study was conducted in the laboratory of Optical Parametric Oscillators group at ICFO-Institut de Ciencies Fotoniques, Barcelona, Spain during author's work in this group. The author thanks the group leader, Prof. Majid Ebrahim-zadeh, for his understanding and granting the permission to publish the work. The author also thanks Prof. Herbert Berk and Ellen Hutchison of the University of Texas at Austin, Prof. Eugene Chudnovsky of the City University of New York (CUNY) and Samuel Markson of Massachusetts Institute of Technology (MIT) for their constant support and encouragement and also their valuble comments and suggestions to improve the manuscript.